# The Phase Dynamics of Earthquakes:
# Implications for Forecasting in Southern California

by

K.F. Tiampo[1], John B. Rundle[2], S. McGinnis[1], W. Klein[3] and S.J. Gross[1]
[1]Colorado Center for Chaos & Complexity
CIRES UCB 216
University of Colorado, Boulder, CO 80309
[2]Department of Physics and Colorado Center for Chaos & Complexity
CIRES UCB 216
University of Colorado, Boulder, CO 80309
and
Exploration Systems Autonomy Division
Jet Propulsion Laboratory
Pasadena, CA 91109

[3]Department of Physics
Boston University, Boston, MA 02215
and
CNLS, Los Alamos National Laboratory
Los Alamos, NM 87545

**ABSTRACT**

We analyze the space-time patterns of earthquake occurrence in southern California using a new method that treats earthquakes as a phase dynamical system. The system state vector is used to obtain a probability measure for current and future earthquake occurrence. Thousands of statistical tests indicate the method has considerable forecast skill. We emphasize that the method is not a model, and there are no unconstrained or free parameters to be determined by fits to training data sets.

Earthquakes strike populated regions without warning, causing great destruction and loss of life [1,2]. Very recent examples include the M ~ 7.6 El Salvador event of January 13, 2001, in which more than 2000 persons died, and the January 26, 2001 India event in which it is expected, at this writing, that more than 20,000 persons have died. Despite the fact that the largest earthquakes produce slip of several meters over fault areas of as much as 50,000 km$^2$, no reliable precursory phenomena have yet been detected [3,4]. It is difficult for most physicists to understand why events of these magnitudes are not preceded by at least some detectable, causal process. Previous efforts to identify the signals premonitory to such events have naturally focused on local regions near the earthquake source [3-6], and various precursory patterns of seismic activity have been proposed [2,5-15]. However, since the hypothesized patterns are localized on the eventual source region, the fact that one must know where the event will occur before these techniques can be applied is a major drawback to their implementation.

Simulations have shown [16,17] that driven threshold systems such as earthquake faults are strongly correlated, mean field systems. Their dynamics can be understood as an example of phase dynamics ("PD"; e.g., ref [16,18,19]). In PD, the evolution in the state of the system between a *base year* $t_b$ and a later time t is characterized by changes in the phase angle of a *state vector* $|S(t_b,t)\rangle$ that represents the change in average activity rate (earthquake occurrence rate) over the interval $(t_b,t)$. Since the mathematical structure associated with PD can be mapped into the mathematics of quantum mechanics, probability measures in a PD system can be readily defined [16,17,19]. These methods are general, and can be used to analyze observed earthquake seismicity in any region.

To summarize our results: Using instrumental seismicity data for southern California, we find considerable support for the theory that real earthquake fault networks are phase dynamical systems. We construct the change in probability $\Delta P(\mathbf{x},t_1,t_2)$ for seismic activity over the time period ($t_1$, $t_2 > t_1$, and find anomalous regions of increased $\Delta P(\mathbf{x},t_1,t_2)$ that are associated with large events, both during the time interval $(t_1,t_2)$, and for future times $t > t_2$. Thousands of statistical Likelihood ratio tests on well-defined null hypotheses indicate that the Phase Dynamical Probability Change (PDPC) index method has considerable forecast skill. Finally, we display a map showing the increase in PDPC index $\Delta P(\mathbf{x},t_1,t_2)$ for southern California as a forecast for the ~10 years following 1999.

The applicability of phase dynamics to earthquake systems follows from physical arguments as well as simulations [16,17]. Within a seismically active region, the well-known Gutenberg-Richter scaling relation [1,2] describes the rate of occurrence r(*m*) of earthquakes over a geographic region with area A, and a time interval T, having magnitude larger than *m*, in terms of two parameters, *a* and *b*:

$$r(m) = 10^a \, 10^{-bm} \qquad (1)$$

Here r(*m*) is number of events per year, per km$^2$ of ground surface area within the scaling regime. Observations indicate that there is a lower cutoff magnitude $m < m_c$ for which $r(m_c)$ = constant, and $r(m) < r(m_c)$. For the earth as whole, the values for the parameters that best fit the data for events within the scaling region [1,2] indicate that a ~ -0.5 and *b* ~ 1.0. The value for *a* differs from that normally quoted in the literature since the number of events is scaled by the total area of the geographic region. For smaller regions such as southern California (figure 1), Japan, and New Zealand, similar values for *a* and *b* are found (see ref [1] for a simple discussion). We conclude that as A and T increase, r(*m*) tends towards a well-defined, world-wide mean for given *m*, with a variance that decreases towards zero. The mean is a direct consequence of the steady rate of energy input from the earth's convective plate tectonic engine.

Now consider a seismically active geographic region having an area A, and a *constant* overall seismic activity rate r(*m*). Construct a spatial coarse-graining by tiling the geographical region with N equal size square sub-regions (boxes), with the i$^{th}$ box centered on a point defined by the position vector in the dual space $\langle \mathbf{x}_i |$. We select the scale of the coarse-graining to resolve the large events that we wish to study. For example, if we wish to study earthquakes of magnitude $m \geq 6$, we select the linear size $\Delta L$ of each box to be $\Delta L \sim 10$ km; similarly, for $m \geq 5$, we select $\Delta L \sim 1$ km, and so forth. The earthquakes occurring within A may be distributed arbitrarily among the N boxes, thus we define $|S(t_b, t)\rangle$ such that the projection onto the position vectors $\langle \mathbf{x}_i |$ is the number of earthquakes $n(\mathbf{x}_i, \Delta t)$ in the box at $\langle \mathbf{x}_i |$ during $\Delta t_b = t - t_b$, per unit time:

$$\langle \mathbf{x}_i | S(t_b, t) \rangle = \langle \mathbf{x}_i | S(t - t_b) \rangle = \frac{n(\mathbf{x}_i, \Delta t_b)}{\Delta t_b} \qquad (2)$$

From the definition of r(*m*):

$$\frac{1}{A} \int_A d\mathbf{x} \, \langle \mathbf{x} | S(t-t_b) \rangle = r(m_c) \tag{3}$$

Since we are interested in analyzing departures from the steady state, we define a *normalized* activity rate vector $|s(t_b,t)\rangle$:

$$\langle \mathbf{x}_i | s(t_b,t) \rangle \equiv \frac{\langle \mathbf{x}_i | S(t_b,t) \rangle - r(m_c)}{\| \langle \mathbf{x}_i | S(t_b,t) \rangle - r(m_c) \|} \tag{4}$$

where $\| \; \|$ represents the L$_2$-norm. $|s(t_b,t)\rangle$ describes the *departure* of the normalized activity rate over the time period ($t_b$, t) from the overall average rate, $r(m_c)$, within A. It is a vector whose *phase angle* is a normalized representation of the average activity rate during $\Delta t_b$ at $\mathbf{x}_i$. The Dirac bra-ket $\langle \mathbf{x}_i | s(t_b,t) \rangle$ also represents a probability amplitude.

Over a time period ($t_1$,$t_2$), the change from $|s(t_b,t_1)\rangle$ to $|s(t_b,t_2)\rangle$ will depend on the choice of base year $t_b$. Since the state vectors $|s(t_b,t)\rangle$ have all been normalized, they should all be treated equally. In defining the change in state $|\Delta s(t_1,t_2)\rangle$, we therefore integrate over all base years up to $t_2$, using equal weighting:

$$\langle \mathbf{x}_i | \Delta s(t_1,t_2) \rangle \equiv \lim_{t_0 \to -\infty} \frac{\int_{t_0}^{t_2} dt_b \, \{ \langle \mathbf{x}_i | s(t_b,t_2) \rangle - \langle \mathbf{x}_i | s(t_b,t_1) \rangle \}}{\int_{t_0}^{t_2} dt_b} \tag{5}$$

Phase dynamics implies that probabilities are squares of state vector amplitudes. Since we are interested in the increase of probability above the time dependent background probability $\mu_B(t_1,t_2)$, we first compute:

$$\mu_B \equiv \frac{1}{A} \int_A d\mathbf{x} \, |\langle \mathbf{x}_i | \Delta s(t_1,t_2) \rangle|^2 \tag{6}$$

The change in probability $\Delta P(\mathbf{x}_i,t_1,t_2)$ for activity above the background is then:

$$\Delta P(\mathbf{x}_i,t_1,t_2) \equiv |\langle \mathbf{x}_i | \Delta s(t_1,t_2) \rangle|^2 - \mu_B(t_1,t_2) \tag{7}$$

Equations (4)-(7), imply that probability $\Delta P(\mathbf{x}_i,t_1,t_2)$ is conserved for any ($t_1$,$t_2$):

$$\int_A d\mathbf{x} \, \Delta P(\mathbf{x}_i,t_1,t_2) = 0 \tag{8}$$

To apply these methods to southern California, we must take the region A large enough so that $A \gg \xi^2$, where $\xi$ is the seismic activity correlation length, estimated to be roughly $\xi \sim$ 200-300 km [20]. Equation (3) can therefore be expected to hold. In southern California the instrumental record of earthquake activity exists only since the year 1932, and is complete only for magnitudes $m \geq 3$. Thus in (5), the lower limit of integration $t_0$ must be replaced by 1932.

We note that there are *no free parameters* to be determined by fits to data. There are certainly decisions to be made, relative to the time period $(t_1,t_2)$ of interest, the geographic extent of the area A, and the minimum size of the large earthquakes to be forecast, the latter determining the spatial coarse-graining scale L. Similar decisions would also be needed for any other method of data analysis, such as taking Fourier transforms, or carrying out a Karhunen-Loeve ("Principal Component") analysis. Once these decisions are made, the method is straightforward and prescribed, and makes use of *all* the data that is consistent with earthquake catalog completeness.

The intensity of seismic activity over the years 1932-1991 is shown in Figure 1, using a scale size of $L \sim 11$ km (0.1° at this latitude), since we are interested in forecasting events of magnitude $m \sim 6$ and larger. Figure 2 is a color-contour plot of $\text{Log}_{10} \Delta P(\mathbf{x}_i,t_1,t_2)$, for locations at which $\Delta P(\mathbf{x}_i,t_1,t_2) > 0$, and where $t_1$ = January 1, 1978, and $t_2$ = December 31, 1991. Inverted triangles on this map indicate large events that occurred during $(t_1,t_2)$, circles represent large events that occurred for later times $t > t_2$. The colored anomalies are associated both with activity during the period from $t_1$ to $t_2$ (inverted triangles), as well as *forecasting* activity that occurs after $t_2$ (circles). *We strongly emphasize that no data were used to construct the colored anomalies in figure 2 from the time after December 31, 1991, a date 6 months prior to the June 27, 1992, m ~ 7.6 Landers earthquake (34° 13' N Latitude, 116° 26' W Longitude).*

Visual inspection of Figure 2 and others like it [21] clearly shows that the method has forecast skill, but rigorous statistical testing is needed for quantification. We carry out such testing using standard methods involving statistical Likelihood ratio tests [21,22]. We used two types of null hypotheses to test the forecast in Figure 2 as a predictor of future activity of large events (circles). 1) We constructed thousands of random earthquake catalogs from the observed catalog by using the same total number of events, but assigning occurrence times from a uniform probability distribution over the years 1932-1991, and distributing them uniformly over the original locations. This procedure produces a Poisson distribution of events in space with an exponential distribution of inter-event times. Randomizing the catalog in this way destroys whatever coherent space-time structure may have existed in the data. Each catalog is used to construct a null hypotheses using a Gaussian density at each coarse-grained point $\mathbf{x}_i$, whose peak value is $\Delta P(\mathbf{x}_i,t_1,t_2) + \mu_B(t_1,t_2)$, the total probability including the background, and whose width is the scale size $L \sim 11$ km. 2) For the second null hypothesis, we used the seismic intensity data in Figure 1 directly as a probability density at $\mathbf{x}_i$, as has been proposed in the literature [23] for the "standard null hypothesis". In figure 3, we show computations of a) Log-Likelihoods for 500 random catalogs of the first type (histogram); b) the Log-Likelihood value for the seismic intensity map in figure 1 (vertical dash-dot line); and c) the Log-Likelihood corresponding to the forecast in Figure 2 (dashed line). Since larger values of Log-Likelihood indicate a more successful hypothesis, we conclude that our method is finding space-time structure in the data corresponding to future large events.

The diffusive, mean field nature of the dynamics, leads to several important predictions: 1) Forecasts such as Figure 2 should convey information for times t approximately in the range: $t_2 + (t_2 - t_1) > t > t_2$; 2) Anomalies of elevated probability having area $\Omega$ should persist for a characteristic time $\tau \propto \Omega^\eta$, where $\eta \sim 1$ [20]; and 3) The dynamics implies that we can compute probabilities using path integral methods [20], an approach that we are currently formulating. Finally, Figure 4 shows a forecast for future large events following 1999, based on changes during the years 1989-1999. This is the most unbiased test possible.

Acknowledgements: This work has been supported by CIRES and NASA student fellowships (KFT and SM); by US DOE grant DE-FG03-95ER14499 (JBR); by NASA grant NAG5-5168 (SJG); and by US DOE grant DE-FG02-95ER14498 (WK).

**References:**
[1] Richter, C.F. *Elementary Seismology*, Freeman, San Francisco (1958).
[2] Scholz, C.H. *The Mechanics of Earthquakes and Faulting*, Cambridge University Press, Cambridge, U.K. (1990).
[3] R.J. Geller, D.D. Jackson, Y.Y. Kagan, and F. Mulargia, *Science*, **275**, 1616-1620 (1997).

[4]  H. Kanamori, pp. 1-19,  in *Earthquake Prediction: An International Review*, ed. D.W. Simpson, II, and P. G. Richards, 1 - 19, AGU, Washington, D.C. (1981).
[5]  L.R. Sykes, B.E. Shaw, and C.H. Scholz,  *Pure Appl. Geophys.*, **155**, 207 (1999).
[6]  I.G. Main,  http://helix.nature.com/debates/earthquake/ equake_frameset.html (1999).
[7]  D.P. Schwartz, *J. Geophys. Res.*, **89**, 5681 (1984).
[8]  W.I. Ellsworth, A.T. Cole, and L.Dietz, *Seis. Res. Lett.*, **69**, 144 (1998).
[9]  K. Mogi, Two kind of seismic gaps. *Pageoph*, **117**, 1172-1186 (1979
[10]  F. Press and C.R. Allen, *J. Geophys. Res.*, **100**, 6421 (1995).
[11] M. Wyss , and R.E. Haberman, *PAGEOPH*, **126**, 333 (1988).
[12] N. Kato, M. Ohtake, and T. Hirasawa, *PAGEOPH*, **150**, 249 (1997).
[13] H. Saleur, C.G. Sammis, and D. Sornette, *J. Geophys. Res.*, **101**, 17661 (1996).
[14] B.E. Shaw, J.M. Carlson, and J. Langer, *J. Geophys. Res.*, **97**, 479 (1992).
[15] C.G. Bufe and D.J. Varnes,  *J. Geophys. Res.*, **98**, 9871 (1993).
[16] J.B. Rundle, W. Klein, K.F. Tiampo, and S.J. Gross, *Phys. Rev. E*, **61**, 2418 (2000).
[17] P.B. Rundle, J.B. Rundle, K.F. Tiampo, J. de sa Martins, S. McGinnis and W. Klein, *Phys. Rev. Lett*., submitted (2000).
[18] J.B. Rundle, W. Klein, S.J. Gross and K.F. Tiampo, pp. 127-146, in *Geocomplexity and the Physics of Earthquakes*, ed. J.B. Rundle, D.L. Turcotte and W. Klein, American Geophysical Union, Washington, DC (2000).
[19] H. Mori and Y. Kuramoto, Dissipative Structures and Chaos, Springer-Verlag, Berlin (1997).
[20] D.D. Bowman, G. Ouillon, C.G. Sammis, A. Sornette and D. Sornette,  *J. Geophys. Res*., **103**, 24359 (1998).
[21] K.F. Tiampo, J.B. Rundle and W. Klein, to be published.
[22] S.J. Gross, and J.B. Rundle, *Geophys. J. Int.*, **133**, 57(1998).
[23] Y.Y. Kagan and D.D. Jackson, Geophys. J. Int., **143**, 438 (2000).

**Figure Captions:**

Figure 1.  Relative seismic intensity in southern California for the period 1932 - December 31, 1991.  Relative intensity is number of earthquakes scaled by the maximum and color scale is linear.

Figure 2.  Color-contour plot of $\text{Log}_{10}\ \Delta P(\mathbf{x}_i, t_1, t_2)$, for locations at which $\Delta P(\mathbf{x}_i, t_1, t_2) > 0$.  Times $t_1$ = January 1, 1978, and $t_2$ = December 31, 1991.  Values are scaled by the maximum and color scale is linear.  Inverted triangles are events that occurred from 1978-1991 with $5 < m < 6$ (smallest triangles); $6 < m < 7$ (intermediate triangles); $7 < m$ (largest triangles).  Circles are events that occurred from 1992 - present, with $5 < m < 6$ (smallest circles); $6 < m < 7$ (intermediate circles); $7 < m$ (largest circles).

Figure 3.  Log-Likelihood plots for 500 random catalogs (histogram); for the seismic intensity map of figure 1 (dash-dot line); and for the forecast in figure 2 (dashed line).  All three methods were scored by the Likelihood test according to how well they forecast the events of magnitude $m > 6$ that occurred on or after January 1, 1992 (circles).

Figure 4.  Color-contour plot of $\text{Log}_{10}\ \Delta P(\mathbf{x}_i, t_1, t_2)$, for locations at which $\Delta P(\mathbf{x}_i, t_1, t_2) > 0$.  Times $t_1$ = January 1, 1989, and $t_2$ = December 31, 1999.  Inverted triangles are events during 1989-1999.

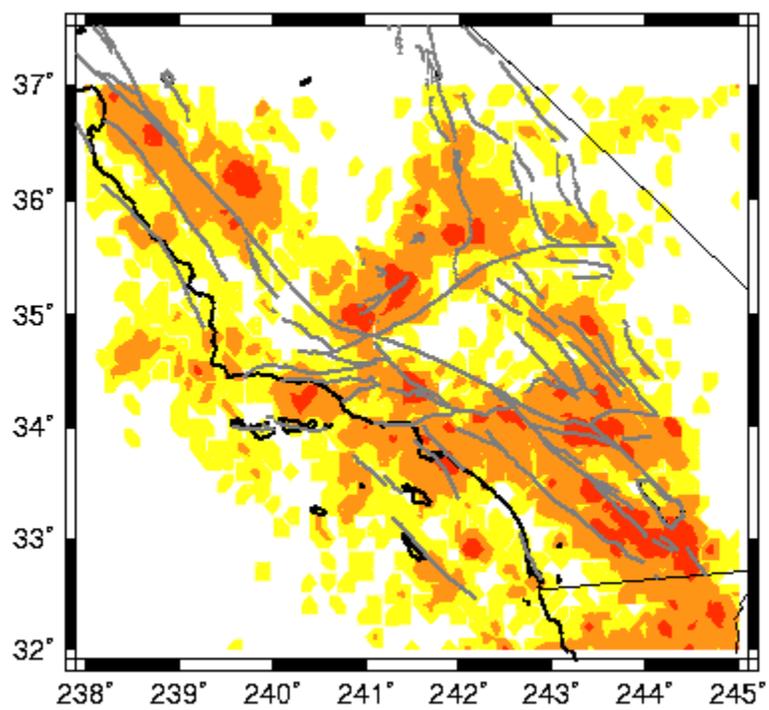

Figure 1

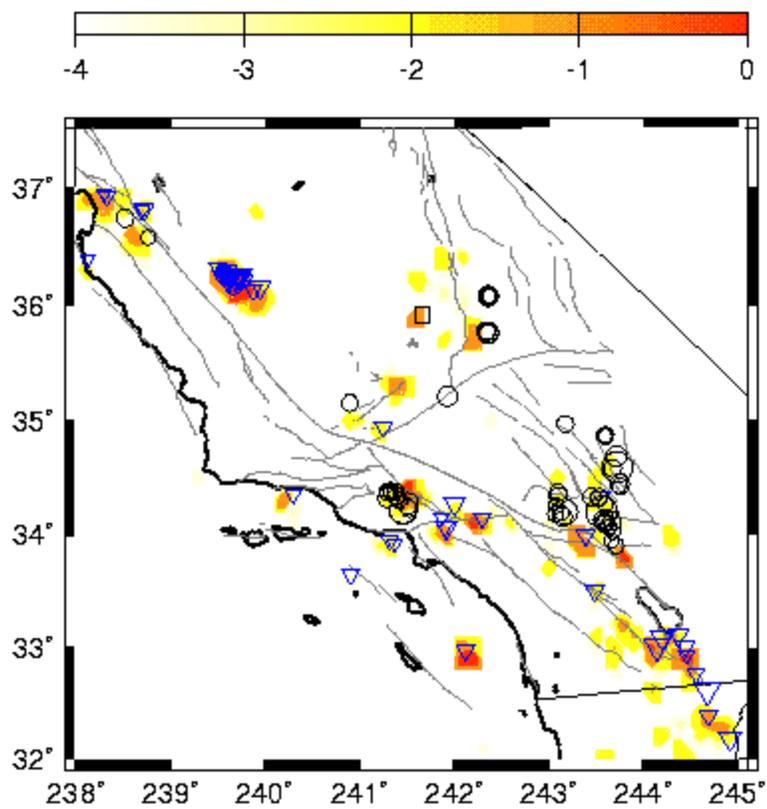

Figure 2

11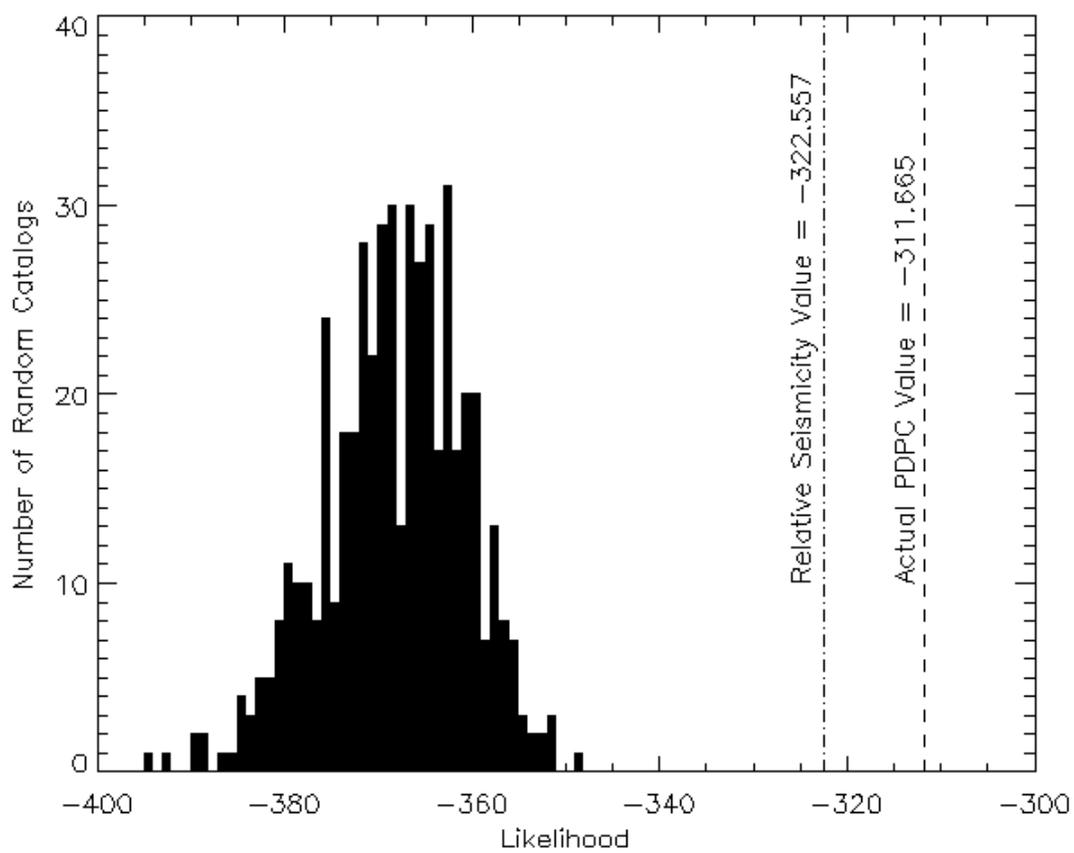

Figure 3

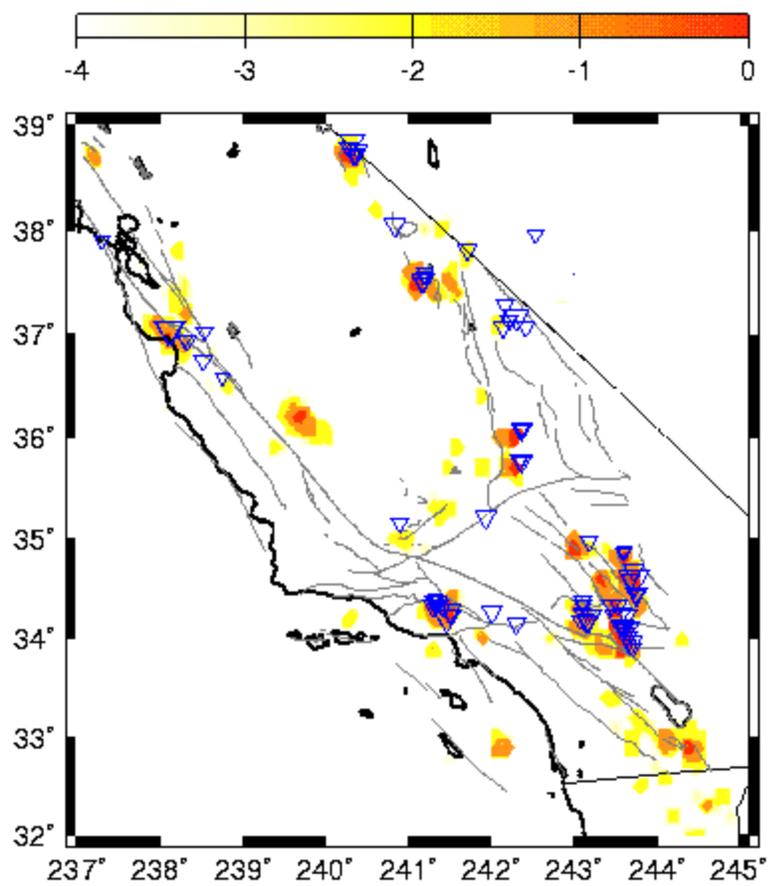

Figure 4